\documentstyle[11pt]{article}
\title{Proper-time methods in the presence of non-constant background fields}
\author{Alan Chodos\\
Center for Theoretical Physics\\
Yale University\\
P.O. Box 208120\\
New Haven, CT 06520-8120}
\begin{document}
\setlength{\baselineskip}{24pt}
\maketitle
\begin{picture}(0,0)(0,0)
\put(295,240){YCTP-P14-94}
\end{picture}
\vspace{-24pt}

\begin{abstract}
\setlength{\baselineskip}{18pt}
A formalism is developed to enable the construction of the effective action and
related quantities in QED for the case of time-varying background electric
fields.  Some examples are studied and evidence is sought for a possible
transition to a phase in which chiral symmetry is spontaneously broken.
\end{abstract}

\medskip

Submitted to the Proceedings of the first G\"{u}rsey Memorial Conference \hfill
\\
(Istanbul, June 6-10, 1994).

\bigskip

The proper-time formalism was used a long time ago by Schwinger  \cite{Schw}
to compute the Green function for an electron propagating in the presence of a
background electromagnetic field.  Although the formalism is general, explicit
evaluation of the propagator, and of the associated effective action, was
possible only for the case of fields uniform in space and constant in time.

Over the intervening decades, attempts have been made \cite{Chan} to compute
the corrections to Schwinger's results for the case of varying fields.  These
take the form of a derivative expansion in the fields, but even the first
non-trivial corrections turn out to be quite unwieldy, and are, moreover,
restricted to fields that do not vary too rapidly (else the higher terms in the
expansion must be included).

More recently, there has been cause for a new look at this problem.  The
motivation is the strange results of the GSI heavy-ion scattering experiments
\cite{Sala}, in which mysterious narrow peaks are seen in the energy spectra of
emitted e+e- pairs.  Among the many theoretical ideas that have been advanced,
I wish to concentrate on one proposed explanation \cite{Cele,DGC1}:  that the
heavy ions create a very strong and rapidly varying electromagnetic field,
which then induces a phase transition in QED to a vacuum in which chiral
symmetry is spontaneously broken.  The observed e+e- peaks are due to the decay
of positronium-like states in the new phase of QED.

To study this possibility, we employ a proper-time representation for the
vaccum expectation value of $\bar{\psi}\psi$, which is an order parameter for
this transition.  This representation is \cite{DGC1}

\begin{equation}
\left\langle\bar{\psi}{\psi} \right\rangle = m\int_{0}^{\infty}
dte^{-m^{2}\tau} U(x,\tau)
\end{equation}

\noindent where

\begin{equation}
U(x,\tau) = tr\left\langle x\mid e^{H\tau}\mid x \right\rangle.
\end{equation}

\noindent Here $\tau$ is the proper-time (continued to imaginary values) and m
is the electron mass.  U is the trace of a quantum-mechanical matrix element
for which the Hamiltonian is $(\gamma\cdot\pi)^2$, with

\begin{equation}
\pi_{\mu} = p_{\mu} - eA_{\mu}(x).
\end{equation}

\noindent The dynamical degrees of freedom are the four coordinates
$x_\mu(\tau)$, and $p_\mu$ are the associated canonical momenta.  $A_\mu(x)$ is
the potential that encodes the background field.  We are working in Euclidean
space, so the $\gamma_\mu$ are Hermitian and H is positive.  The trace in eq.
(2) is over the indices carried by the $\gamma$ matrices.  For later reference,
note that H possesses a quantum-mechanical supersymmetry \cite{DGC1}, generated
by the charges $Q_{\pm}={1 \over 2}  (1_{\pm} \gamma_5)(\gamma\cdot\pi)$ with

\begin{equation}
\{Q_{+}, Q_{-}\} = H .
\end{equation}

The proper-time expression for $\left\langle \bar{\psi}\psi \right\rangle$
incorporates correctly all the effects of the background field, but completely
neglects the role of the dynamical photons.  In using eq. (1), one hopes that
these photons will not affect the presence or absence of a phase transition
induced by the background field.

The signal for the spontaneous breakdown of chiral symmetry is that the limit
$m \rightarrow 0$ of $\left\langle \bar{\psi}\psi \right\rangle$ should not
vanish.  Because of the explicit factor of $m$ in eq. (1), one requires the
integral to diverge as $m \rightarrow 0$.  In fact, one easily sees \cite{Bank}
that if the large-$\tau$ behavior of $U(x,\tau)$ is $\tau^{- {1 \over 2}}$,
$\left\langle \bar{\psi}\psi \right\rangle$ will remain finite and non-zero.
If the falloff is more rapid, $\left\langle \bar{\psi}\psi \right\rangle$ will
vanish, indicating that there is no spontaneous chiral symmetry breaking.

For a free fermion, $U \sim1/ \tau^2$ so that $\left\langle \bar{\psi}\psi
\right\rangle\rightarrow 0$ as expected.  For constant $F_{\mu\nu}$, one finds
from the analytic continuation of Schwinger's results that $U(x,\tau)$ is a
function of the two Euclidean invariants $F = {1 \over 2} (\vec{E}^{2} +
\vec{B}^{2})$ and $G = \vec{E} \cdot \vec{B}$ .  If $G = 0$, $F \neq 0$, $U
\sim 1/\tau$ and $\left\langle \bar{\psi}\psi \right\rangle\rightarrow0$.  If
$G \neq 0$, then $U \sim const.$, which indicates that $\left\langle
\bar{\psi}\psi \right\rangle\rightarrow \infty$  as $m=\rightarrow 0$.  This
behavior is not to be interpreted as spontaneous chiral symmetry breaking,
however, since there is an anomaly when $G \neq 0$ that explicitly breaks
chiral symmetry.  In what follows, we shall look at cases where  $\vec{E} \neq
0$,  $\vec{B} = 0$, so the anomaly is absent.

In a recent paper, Caldi and Vafaeisefat \cite{DGC2} have computed $U(x,\tau)$
numerically using Monte Carlo simulation techniques.  For this purpose, it is
convenient to recast $U(x,\tau)$ as a path integral:

\begin{equation}
U(x,\tau) = tr \int Dx_{\mu}Te^{-S},
\end{equation}

\begin{equation}
S = \int_{0}^{\tau} dt^\prime L(x,\dot{x})  ,
\end{equation}

$L(x,\dot{x}) = {1 \over 4} \dot{x}_{\mu}\dot{x}_{\mu} + ie
\dot{x}_{\mu}A_{\mu}(x) - {e \over 2} \sigma_{\mu\nu} F_{\mu\nu}(x)$.  Note the
following peculiarities:  (i) L is complex (this is a consequence of having
continued to imaginary proper-time); and (ii) L is matrix valued.  The symbol T
in eq. (5) denotes $\tau$-ordering.

Caldi and Vafaeisefat look initially at background electric field
configurations pointing in one direction only, for which the magnitude varies
in time and in the one spatial variable.  In particular, they consider

\begin{equation}
\vec{E} = (f(x,t), 0,0)
\end{equation}

\noindent with

\begin{equation}
f(x,t) = eE[cosh^{2}(x/W_{s})]^{-1} exp(-t^{2}/2 W_{t}^{2}).
\end{equation}

\noindent They find, for $W_{s} = W_{t}^{-1} = 3$ (in units where $eE = 1$)
that at suitably chosen values for $x_{\mu}, U(x,\tau)$ exhibits the desired
$\tau^{-1/2}$ falloff.  Although the computations are complicated, their method
gains credence from the following observations:  they obtain agreement with
Schwinger's analytic results for the case $\vec{E} = const.$, and, when $W_{s}$
and $W_{t}$ are taken much larger or much smaller than the above values, the
chiral symmetry breaking goes away.  This is reasonable because, for large
$W_{s}$ and $W_{t}$ the configuration approximates a constant field, for which
chiral symmetry is unbroken, whereas for small $W_{s}$ and $W_{t}$ the field is
varying so rapidly that the vacuum does not have time to realign (i.e., the
"sudden approximation" is valid).  In later work \cite{DGC3}, Caldi and
Vafaeisefat have studied more realistic configurations involving all the
spatial variables, and they continue to see chiral symmetry breaking for
suitable values of the parameters.

Even assuming the utter reliability of these results, one is still left with
virtually no intuition or insight concerning the mechanism whereby chiral
symmetry is broken.  It is therefore of interest to explore these questions in
a more analytic fashion.  To this end, we look at a configuration even simpler
than that chosen by Caldi and Vafaeisefat, to wit one in which the electric
field depends only on time (which is, of course, Euclidean time, and which we
call $x_{0}$):

\begin{equation}
\vec{E} = (f(x_{0}), 0,0)  .
\end{equation}

\noindent We take the associated vector potential to be

\begin{equation}
A_{0} = - x_{1} f(x_{0})
\end{equation}

\noindent and

\begin{equation}
\vec{A} = 0  .
\end{equation}

The Lagrangian then reduces to:

\begin{equation}
L = {1 \over 4} (\dot{x}_{0}^{2} + \dot{x}_{1}^{2}) + ie \dot{x}_{0}x_{1}
f(x_{0}) + e \sigma_{01} f(x_{0}) + {1 \over 4}  (\dot{x}_{2}^{2} +
\dot{x}_{3}^{2})  .
\end{equation}

\noindent Note the following:

\noindent (a) For proper normalization (i.e. to obtain the known result when
$f(x_{0}) = 0$) we must take

\begin{equation}
\int Dye^{- {1 \over 4}\int_{0}^{\tau}\dot{y}^{2}dt} = {1 \over \sqrt{4\pi
\tau}}  ;
\end{equation}

\noindent (b)  The $x_{2}$ and $x_{3}$ integrals are then trivial, yielding a
factor of ${1 \over 4\pi\tau}$ ;

\noindent (c) The $x_{1}$ integral is almost trivial, since it can be reduced
to a Gaussian by a shift in $x_{1}$;

\noindent (d) the $\tau$-ordering is superfluous because there is only one
non-trivial matrix, $\sigma_{01}$.  Furthermore, the trace is reduced to a
summation over the eigenvalues of $\sigma_{01}$, i.e. to the operation
$2\sum_{\sigma = \pm1}$ (where we have abbreviated $\sigma_{01}$ by $\sigma$).

\noindent (e)  The path integral must be evaluated subject to the boundary
condition $x_{\mu}(0) = x_{\mu}(\tau) = x_{\mu}$, where $x_{\mu}$ is the
argument of  $\bar{\psi}\psi$.

Making use of standard manipulations, one obtains

\begin{equation}
U(x,\tau) = {1 \over 4(\pi \tau)^{3/2}} \sum_{\sigma = \pm 1} \int Dx_{0}e^{-
{1 \over 4} \int_{0}^{\tau}dt^\prime \dot{x}_{0}^{2}} e^{-e^2 \tau [<F^2 > -
<F>^2]} e^{-e \sigma \int_{0}^{\tau} d \tau \prime f}.
\end{equation}

\noindent Here $F$ is defined by ${dF \over dx_0} = f(x_0)$, and for any
function $\Phi (\tau)$, we have defined $\left\langle \Phi  \right\rangle = {1
\over \tau} \int_{0}^{\tau}d\tau \prime \Phi (\tau^{\prime})$.  Note that $U$
is invariant under $F(x_0) \rightarrow F(x_0) + C$, as it must be.  Also note
that the $\left\langle F \right\rangle^2$ term is a non-local interaction.

We can re-express $U(x,\tau)$ in terms of a local action at the cost of
introducing an ordinary integral over a parameter $\lambda$.  Using

\begin{equation}
e^{\beta ^2/4 \alpha} = \sqrt{{\alpha \over \pi}} \int_{- \infty}^{\infty} d
\lambda e^{- \alpha \lambda^2 + \beta \lambda}
\end{equation}

\noindent with $\alpha =  \tau$ and $\beta = 2e\tau \left\langle F
\right\rangle$, we have

\begin{equation}
U(x,\tau) = {1 \over 4\pi^2 \tau} \sum_{\sigma = \pm 1} \int D \lambda \int_{-
\infty}^{\infty} Dx_0 e^ {-\int_{0}^{\tau} d \tau \prime L_\sigma}.
\end{equation}

\noindent where

\begin{equation}
L_\sigma = {1 \over 4} \dot{x}_0^2 + (eF - \lambda)^2 + e\sigma f .
\end{equation}

\noindent Thus we are summing and integrating over a family of one-dimensional
quantum-mechanical models defined by the Hamiltonians

\begin{equation}
H_\sigma = p^2 + V_\sigma (x_0)    ,
\end{equation}

\begin{equation}
V_\sigma = [eF(x_0) - \lambda]^2 + e\sigma f(x_0)   .
\end{equation}

\noindent Here we see that $H_\pm$ have the standard form \cite{Witt}
characteristic of quantum-mechanical supersymmetry,

\begin{equation}
H_\pm = p^2 + W^2 \pm dW/dx_0  ,
\end{equation}

\noindent with

\begin{equation}
W = eF - \lambda   .
\end{equation}

For the purpose of quantitative analysis, it is convenient to re-express
$U(x,\tau)$ as

\begin{equation}
U(x,\tau) = {1 \over 4\pi^2 \tau} \sum_{\sigma = \pm 1} \int_{-
\infty}^{\infty} d \lambda < x \mid e^{-H_{\sigma} \tau} \mid x >  ,
\end{equation}

\noindent and then to insert this in eq. (1), and perform the $\tau$ integral
after division by $m$ and differentiation with respect to $m^2$.  One obtains

\begin{equation}
I(m) = - {\partial \over \partial m^2} [{<\bar{\psi}\psi (x)> \over m}] = {1
\over 4\pi^2} \sum_{\sigma} \int d \lambda <x \mid{1 \over H_{\sigma} +m^2}
\mid x >   .
\end{equation}

\noindent Therefore, we wish to compute the Green function $G_\sigma
(x,x^\prime) = <x\mid{1 \over H_{\sigma} +m^2} \mid x^\prime >$, which obeys
the equation

\begin{equation}
[- {\partial^2 \over \partial x^2} + V_\sigma (x) + m^2 ] G(x,x^\prime ) =
\delta(x - x^\prime )   .
\end{equation}

\noindent A standard expression for $G$ is

\begin{equation}
G(x,x^{\prime}) = \psi_{\scriptscriptstyle >} (x) \psi_{\scriptscriptstyle <}
(x^{\prime}) \theta (x - x^{\prime}) + \psi_{\scriptscriptstyle >} (x^{\prime})
\psi_{\scriptscriptstyle <} (x) \theta (x^{\prime} - x)
\end{equation}

\noindent where each $\psi$ obeys the homogeneous equation

\begin{equation}
[ - {\partial ^2 \over \partial x^2} + V_\sigma + m^2] \psi = 0   ,
\end{equation}

\noindent subject to the boundary condition that $\psi_{\scriptscriptstyle >}
(\psi_{\scriptscriptstyle <})$ is well-behaved as $x \rightarrow \infty (x
\rightarrow - \infty)$, and where the Wronskian condition

\begin{equation}
\psi_{\scriptscriptstyle >} {\partial \psi_{\scriptscriptstyle <} \over
\partial x} - \psi_{\scriptscriptstyle <} {\partial \psi_{\scriptscriptstyle >}
\over \partial x} = 1
\end{equation}

\noindent is imposed.  We then have

\begin{equation}
I(m) = {1 \over 4\pi^2} \int_{- \infty}^{\infty} d\lambda \sum_{\sigma = \pm 1}
\psi_{\scriptscriptstyle >}^{\sigma}(x) \psi_{\scriptscriptstyle <}^\sigma (x)
{}.
\end{equation}

\noindent Our computational strategy is to choose f(x) so that $\psi_>$ and
$\psi_<$ can be computed explicitly \cite{AGI,CGK}, to insert them in eq. (28),
and to determine therefrom the behavior of $I(m)$ as $m \rightarrow 0$.  If
$\left\langle \bar{\psi} \psi \right\rangle$ indeed tends to a finite, non-zero
value, we expect $I(m) \sim 1/m^3$.  Any less singular behavior will be
evidence that $\left\langle \bar{\psi} \psi \right\rangle$ is tending to zero.

Actually, without any further computation, we can infer from eq. (23) that
$I(m)$ will probably behave as $1/m^2$, provided there is some range of
$\lambda$ for which supersymmetry is unbroken.  Under these circumstances, one
of $H_\pm$ will have zero as its lowest eigenvalue, and hence the matrix
element expanded in energy eigenstates will have a term that goes as $1/m^2$.
For the range of $\lambda$ (if any) for which supersymmetry is broken, the
situation is worse:  Both $H_+$ and $H_-$ will have strictly positive ground
state energies, so the contribution to $I(m)$ will be non-singular.  It is hard
to imagine a system for which the desired $1/m^3$ singularity might appear.

As an illustrative example, we can choose

\begin{equation}
eF(x) = \gamma tanh \beta x  .
\end{equation}

\noindent This yields a model that is exactly solvable quantum mechanically
\cite{CGK}.  One finds that supersymmetry is unbroken for $\mid\lambda /
\gamma\mid < 1$, and broken when $\mid\lambda / \gamma\mid >1$.  For $\lambda =
\pm \gamma$, the zero-energy eigenstate is not isolated but sits at the bottom
of a continuous spectrum.  Some of the energy levels of this model as a
function of $\lambda$ are illustrated in Figure 1.  For any values of $\gamma$
and $\lambda$ one can solve for $\psi_{\scriptscriptstyle >}$ and
$\psi_{\scriptscriptstyle <}$ explicitly in terms of hypergeometric functions
$_2F_1$.  We do not reproduce the formulas here, since they are complicated and
not particularly illuminating.  When inserted into eq. (28), they yield the
expected $1/m^2$ behavior, i.e. no evidence for chiral symmetry breaking.

It is possible to study other exactly solvable quantum mechanical models as
well.  Examples are available for which supersymmetry is unbroken for all
$\lambda$ and there are others for which supersymmetry is broken for all
$\lambda$ except $\lambda = 0$.  In all the cases we have examined, the
singularity at small $m$ of $I(m)$ is no worse than $1/m^2$.

This result is not in conflict, of course, with the numerical work of Caldi and
Vafaeisefat, since their field configurations depend on at least two variables.
 In deciding how to proceed, one can think of a number of possibilities:  (i)
extend the search among the one-dimensional models in the hope that an as yet
undiscovered class will yield the sought-for $1/m^3$ behavior; (ii) introduce
new analytic techniques that will enable one to study the two-variable case.
This will permit direct comparison with the Monte Carlo results; (iii) find a
way to extract the small $m$ behavior of $I(m)$ (or equivalently the large t
behavior of $U(x,\tau)$) without first having to compute the full functional
forms of $I(m)$ or $U(x,\tau)$.  This would lead to enormous simplifications
not only of the analytic work but also of the Monte Carlo calculations, where
the large $\tau$ behavior is extracted by computing $U(x,\tau)$ for several
values of  $\tau$ and finding the slope of the best-fitting straight line on a
log-log plot.

As new data from GSI and Argonne are reported, one expects the relevance of the
ideas upon which the present work is based either to wax or to wane.  If the
former, it will be interesting to see whether new insight can in fact be gained
about the mechanism whereby time- and space-varying background fields induce a
chiral phase transition in QED.

\noindent\medskip{\bf Acknowledgments}

I am especially grateful to Daniel Caldi, Andras Kaiser, David Owen and Saeid
Vafaeisefat, each of whom contributed substantially to various parts of the
work described in this paper.  I am also grateful for illuminating
conversations with Charles Sommerfield.  I am pleased that this work touches on
the subject of exactly solvable quantum mechanical models, because this is a
subject in which Feza was interested and to which he made significant
contributions in collaboration with Franco Iachello and Yoram Alhassid.
Finally I wish to express my deepest thanks to Meral Serdaroglu and the rest of
the organizers at Bogazici University, to whom the success of the first
G\"{u}rsey Memorial Conference is in large measure due.  The research discussed
in this paper was supported in part by the U.S. Department of Energy grant
DEFG0292ER40704.

\noindent Figure Caption

\noindent The first two energy levels of the  $F = \gamma tanh \beta x$  model
discussed  in the text.   $E_0 = 0$   for  $\lambda < \lambda_0 = \gamma$, and
$E_0 = E_c = (\gamma - \lambda)^2$ for $\lambda > \gamma$.  $E_c$ is the energy
at which the continuous spectrum begins.  The first excited bound state exists
for $\lambda < \lambda_1 = \hfill \\
(\gamma - \beta)^{2}/\gamma$ (provided $\beta < \gamma$), and is given
explicitly by $E_1 = \hfill \\
\beta (2\gamma - \beta)(1 - \lambda^2/(\gamma - \beta)^2)$.  The diagram is
symmetric for $\lambda \rightarrow - \lambda$.

\noindent Figure available by request (chodos@yalph2.physics.yale.edu).

\end{document}